\documentclass[12pt]{article}
\usepackage[dvips]{graphicx,psfrag}
\usepackage{amssymb}
\input epsf
\newcommand{\be}{\begin{equation}}
\newcommand{\ee}{\end{equation}}
\newcommand{\bea}{\begin{eqnarray}}
\newcommand{\eea}{\end{eqnarray}}

\begin{document}
\begin{titlepage}
\begin{center}
{\large\bf
GAUSS-BONNET BLACK HOLES AT THE LHC~:  BEYOND  THE  DIMENSIONALITY  OF
SPACE}
\vskip 2 cm
{\bf A. Barrau${}^{1,*}$, J. Grain${}^1$, S. Alexeyev${}^{1,2}$}
\vskip 0.4cm
 ${}^1$Laboratory for Subatomic Physics and Cosmology\\
 Joseph Fourier University, CNRS-IN2P3 \\
 53, avenue des Martyrs, 38026 Grenoble cedex, France
\vskip 0.7cm
 ${}^2$Sternberg Astronomical Institute \\
 Lomonosov Moscow State University \\
 Universitetsky Prospect, 13, Moscow 119992, Russia

\end{center}
%\date{\today}
\vskip .3cm
\centerline{${}^*$ corresponding author, e-mail : \tt  Aurelien.Barrau@cern.ch}

\vskip 1cm
\small
\begin{center}
{\bf Abstract}
\end{center}

\begin{quote}
The Gauss - Bonnet invariant is one of the most promising candidates for
a quadratic curvature  correction to the Einstein action in expansions
of  supersymmetric  string theory. We study the  evaporation  of  such
Schwarzschild - Gauss - Bonnet  black  holes which could be formed  at
future colliders if the  Planck scale is of order a TeV,  as predicted
by  some  modern  brane  world  models.  We  show   that,  beyond  the
dimensionality of space, the corresponding coupling  constant could be
measured by the LHC. This opens new windows  for physics investigation
in spite of the possible  screening  of microphysics due to the  event
horizon.
\end{quote}
\normalsize
\date{\today}

PACS Numbers: 04.70.Dy (Quantum aspects of black holes),
              11.25.-w (Strings and branes),
              13.90.+i (phenomenology of elementary particles)

\end{titlepage}

\section{Introduction}

It has recently been  pointed out that black holes could be  formed at
future colliders if the Planck scale is of order a TeV, as is the case
in some  extra-dimension  scenarios  \cite{dimo,gidd}.  This  idea has
driven a considerable amount of interest (see {\it e.g.} \cite{idea}).
The same phenomenon could also occur due to  ultrahigh energy neutrino
interactions in the atmosphere \cite{ancho}. Most  works consider that
those black holes could be described by the D-dimensional ($D  \ge 5$)
generalized Schwarzschild or Kerr metrics  \cite{myers}.
The aim of this paper is to study the experimental consequences of the
existence of the Gauss - Bonnet term (as a step toward quantum gravity) if it
is included in the D-dimensional action. This approach  should be more
general  and  relies  on  a  real  expansion of supersymmetric  string
theory. In Section 2, the basics of black hole formation at
colliders and the  related cross sections are reminded. The details of
the multi-dimensional Gauss - Bonnet black hole solutions and their thermodynamical
properties are  given in Section 3. The flux  computation and the main
analytical formulae  are explained in  Section 4. It is shown
in Section  5 that the  Gauss - Bonnet (string) coupling constant can be
measured in most cases, together with  the  dimensionality  of  space.
Finally, some  possible  consequences  and  developments,
especially with an additional cosmological constant, are discussed.

\section{Black hole formation at colliders}

The "large extra  dimensions" scenario \cite{large} is a very exciting
way to address geometrically the  hierarchy  problem  (among  others),
allowing only the gravity to propagate in the bulk. The Gauss law relates
the Planck  scale of the  effective 4D low-energy theory $M_{Pl}$ with
the  fundamental  Planck   scale  $M_D$  through  the  volume  of  the
compactified dimensions, $V_{D-4}$, via:
$$M_{D}=\left(\frac{M_{Pl}^2}{V_{D-4}}\right)^{\frac{1}{D-2}}.$$
It is thus possible  to set $M_D\sim  TeV$ without being  in  contradiction
with any currently available
experimental data. This translates into radii values 
between a  fraction of a
millimeter and a  few  Fermi for  the  compactification radius of  the
extra dimensions (assumed to  be of same size and flat, {\it  i.e.} of
toroidal shape). Furthermore, such a small value for the Planck energy
can be naturally expected to minimize the difference  between the weak
and Planck  scales, as motivated by the construction of this approach.
In such a scenario, at sub-weak energies, the Standard Model (SM) fields 
must be localized to a 4-dimensional manifold of weak scale "thickness" 
in the extra dimensions. As shown in \cite{large}, as an example based 
on a dynamical assumption with D=6, it is possible to build such a 
SM field localization. This is however the non-trivial task of those models.

Another  important  way for realizing TeV scale  gravity  arises  from
properties   of   warped   extra-dimensional   geometries   used    in
Randall-Sundrum scenarios \cite{rs}. If the  warp  factor  is small in
the vicinity of the standard model brane, particle masses can take TeV
values, thereby  giving rise to a large hierarchy  between the TeV and
conventional  Planck  scales  \cite{gidd,katz}.  Strong  gravitational
effects  are  therefore  also  expected  in   high  energy  scattering
processes on the brane.

In those  frameworks, black holes could be formed  by the Large Hadron
Collider  (LHC).  Two partons with a center-of-mass energy  $\sqrt{s}$
moving in  opposite directions with  an impact parameter less than the
horizon  radius  $r_+$  should  form  a  black  hole  of mass  $M  
\approx \sqrt{s}$ with a cross  section expected  to  be of  order
$\sigma \approx  \pi r_+^2$. Thoses values are in fact approximations as
the black hole mass will be only a fraction of the center-of-mass energy 
whose exact value depends on the dimensionality of the spacetime and 
the angular momentum of the produced black hole \cite{supp,supp2} .
Furthermore, suppression effects in the cross section should be considered  
and are taken into account in the section 5 of this paper. Although the
accurate values are not yet known, a semiclassical analysis
of quantum black hole formation is now being constructed and the existence
of a closed trapped surface in the collision geometry of relativistic particles
is demonstrated.
To compute  the real probability  to form
black holes at the LHC, it is necessary to take into account that only
a fraction  of the total center-of-mass energy is  carried out by each
parton  and  to   convolve  the  previous  estimate  with  the  parton
luminosity  \cite{dimo}.  Many  clear   experimental   signatures  are
expected \cite{gidd}, in particular very high multiplicity events with
a large  fraction of the  beam energy converted into transverse energy
with a growing  cross  section. Depending on the  value  of the Planck
scale, up to approximately a billion black holes could be  produced at
the LHC.

\section{Schwarzschild - Gauss - Bonnet black holes}

\begin{figure}
[htbp]
$$
\epsfxsize=9cm
\epsfysize=7cm
\epsfbox{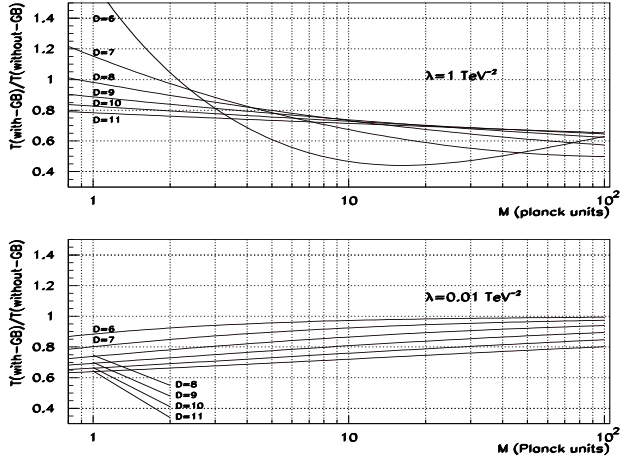}
$$
\caption{Ratio of the  temperatures  with and without the Gauss - Bonnet
term for $D=6,7,8,9,10,11$ (from up to bottom in the low  mass region)
as  a  function  of  mass   with   $\lambda=1$~TeV$^{-2}$   (up)   and
$\lambda=0.01$~TeV$^{-2}$ (down).}
\label{tbh}
\end{figure}

The classical Einstein theory can be considered as the weak  field and
low energy  limit of a  some quantum  gravity model which  is not  yet
built. The curvature expansion of string gravity therefore provides an
interesting step in  the modelling of a quasiclassical approximation of
quantum gravity. As pointed out  in  \cite{boulw},  among higher order
curvature corrections to the general relativity  action, the quadratic
term  is  especially  important as it is the leading one and as it can
affect the graviton excitation spectrum near flat space.  If, like the
string itself, its slope expansion is to be ghost free,  the quadratic
term {\it must} be the Gauss - Bonnet combination~:
$L_{GB} = R_{\mu\nu\alpha\beta} R^{\mu\nu\alpha\beta} -
4 R_{\alpha\beta} R^{\alpha\beta} + R^2$.
Furthermore, this  term  is  naturally  generated  in heterotic string
theories \cite{zw} and makes possible the localization of the graviton
zero-mode on the brane \cite{mavro}. It has been  successfully used in
cosmology, especially  to  address  the  cosmological constant problem
(see {\it e.g.} \cite{cosmo} and references therein) and in black hole
physics, especially to address the endpoint of the Hawking evaporation
problem (see {\it e.g.} \cite{bh} and references therein). We consider
here black holes described by such an action~:
$$S=\frac{1}{16\pi G}\int d^Dx\sqrt{-g}\left\{ R + \lambda (
R_{\mu\nu\alpha\beta} R^{\mu\nu\alpha\beta}
- 4 R_{\alpha\beta} R^{\alpha\beta} + R^2) + \ldots
\right\},$$
where $\lambda$ is the Gauss - Bonnet coupling constant. The measurement
of this $\lambda$ term would  allow  an important step forward in  the
understanding of the  ultimate  gravity theory. Following \cite {cai},
we assume the metric to be of the following form~:
$$ds^2=-e^{2\nu}dt^2+e^{2\alpha}dr^2+r^2h_{ij}dx^idx^j$$
where $\nu$ and $\alpha$ are  functions  of $r$ only and $h_{ij}  dx^i
dx^j$  represents   the   line   element   of   a  $(D-2)$-dimensional
hypersurface with  constant  curvature $(D-2) (D-3)$. The substitution
of this metric  into the action  \cite{boulw} leads to  the  following
solutions~:
$$e^{2\nu}=e^{-2\alpha}=1+\frac{r^2}{2\lambda(D-3)(D-4)}\times$$
$$\left(1\pm
\sqrt{1+\frac{32\pi^{\frac{3-D}{2}}G\lambda
(D-3)(D-4)M\Gamma(\frac{D-1}{2})}{(D-2)r^{D-1}}} \right).$$
The mass  of the black hole can then  be expressed \cite{boulw,cai} in
terms of the horizon radius $r_+$,
$$M = \frac{(D-2) \pi^{\frac{D-1} {2}} r_+^{D-3}} {8\pi
G\Gamma\left( \frac{D-1}{2} \right)}
\left(1 + \frac{\lambda (D-3) (D-4)}{r_+^2} \right)$$
where
$\Gamma$ stands for the Gamma function. The temperature is obtained by
the  usual  requirement  that  no  conical  singularity appears at  the
horizon in the euclidean sector of the hole solution,
$$T_{BH}
=\frac{1}{4\pi}
(e^{-2\alpha})'\mid_{r = r_+}
=
\frac{(D-3) r_+^2 + (D-5)(D-4)(D-3) \lambda} {4\pi
r_+\left( r_+^2 + 2\lambda(D-4) (D-3) \right)}.$$
In  the  case  $D=5$,  those  black  holes have  a  singular  behavior
\cite{cai}  and,  depending  on  the  value  of $\lambda$, can  become
thermodynamically unstable or  form stable relics. For $D>5$, which is
the only relevant hypothesis for this study (as $D=5$ would  alter the
solar  system  dynamics  if  the  Planck  scale  is  expected  to  lie
$\sim$TeV),  a  quantitatively  different   evaporation   scenario  is
expected. Figure \ref{tbh} shows the  ratio  of  the temperatures with
and without the  Gauss - Bonnet  term for  different  values of $D$  and
$\lambda$. It should be pointed  out  that  the non-monotonic behavior
makes an unambiguous  measurement quite difficult and requires to take
advantage of  the full dynamics  of the evaporation. The next sections
focus  on  this  point  to   investigate   the   $\lambda$   parameter
reconstruction.

\section{Flux computation}

\begin{figure}
[htbp]
$$
\epsfxsize=15cm
\epsfysize=8.5cm
\epsfbox{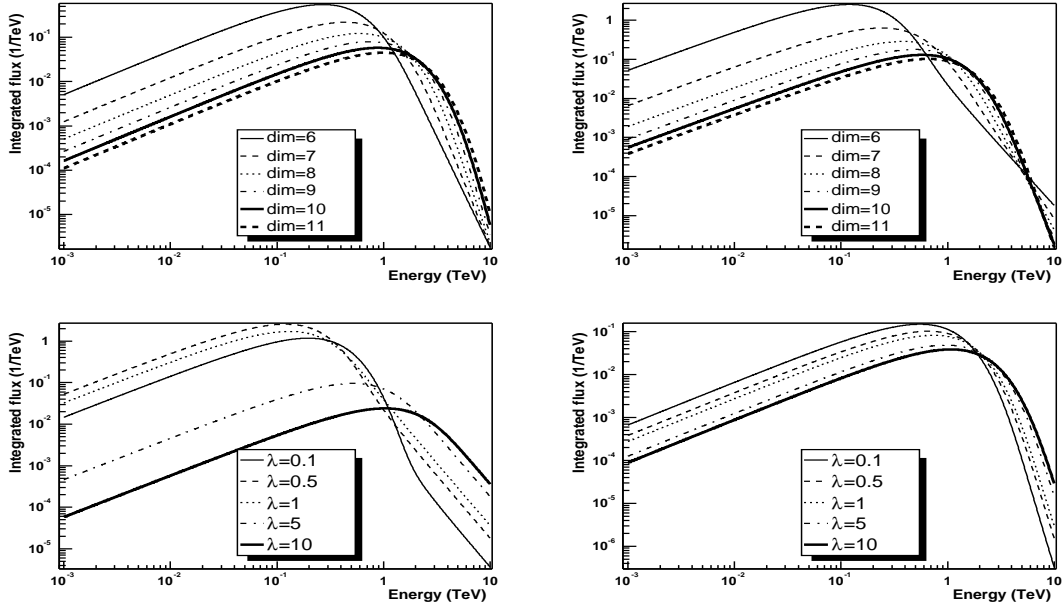}
$$
\caption{Integrated flux as  a  function of  the  total energy of  the
emitted quanta for an initial black hole mass $M=10$~TeV. Upper  left :
$\lambda=0$,  $D=6,  7, 8,  9,  10,  11$.  Upper  right  : $\lambda=0,
5$~TeV$^{-2}$,  $D=6,   7,  8,  9,   10,  11$.  Lower  left  :  $D=6$,
$\lambda=0.1,  0.5,   1,5,   10$~TeV$^{-2}$.  Lower  right  :  $D=11$,
$\lambda=0.1, 0.5, 1, 5, 10$~TeV$^{-2}$.}
\label{spec}
\end{figure}

Using the  high-energy  limit  of  multi-dimensional grey-body factors
\cite{kanti}, the spectrum per unit of  time $t$ and of energy $Q$ can
be written, for each degree of  freedom, for particles of type $i$ and
spin $s$ as:
$$\frac{{\rm d}^2 N_i}{{\rm d}Q{\rm d} t} = \frac{4 \pi^2\left(
\frac{D-1}{2}
\right)^{\frac{2} {D-3}} \left( \frac{D-1}{D-3} \right)
r_+^2Q^2}{e^{\frac{Q}{T_{BH}}}-(-1)^{2s}}.$$
This is  an                 
approximation as modifications might arise when the exact values               
of the greybody factors are taken into account due to their             
dependence, in the low energy regime, on both the dimensionality of the              
spacetime and on the spin of the emitted particle. Fortunately,
as demonstrated in the 4-dimensional case \cite{barrau}, the
{\it pseudo-oscillating} behaviour induces compensations that makes
the differences probably quantitatively quite small.      
As shown in the previous section, as long as $D>5$, the horizon radius
$r_+$ cannot  be explicitly given as a  function of  the mass and,  to
compute the experimental integral spectrum ${\rm  d}N_i/{\rm d}Q$, the
following change of variable is convenient~:
$$\frac{{\rm d}N_i}{{\rm d}Q}
= \int_{r_{init +}}^0\frac{1} {\frac{{\rm d}M}{{\rm
d}t}}\frac{{\rm d}M}{{\rm d}r_+} \frac{{\rm d}^2N_i} {{\rm d}
Q{\rm d}t}{\rm d}r_+$$
where
$$\frac{{\rm d}M}{{\rm d}r_+}
= \frac{(D-2)\pi^{\frac{D-1}{2}}r_+^{D-6}}{8\pi G \Gamma
( \frac{D-1}{2} )}\left[ (D-3)r_+^2+(D-5)(D-4)(D-3)\lambda \right]$$
and
$$\frac{{\rm d}M}{{\rm d}t}=-\frac{4\pi ^6}{15}
\left( \frac{D-1}{2}
\right)^{\frac{2}{D-3}}\left( \frac{D-1}{D-3} \right)
r_+^2 T_{BH}^4\left[\frac{7}{8}N_f+N_b\right],$$
$N_f$  and $N_b$  being  the total fermionic  and  bosonic degrees  of
freedom. The mean number of emitted particle can then be written as
$$N_{tot}
= \frac{15(D-2)\pi^{\frac{D-9}{2}}\zeta(3)}{\Gamma(\frac{D-1}{2})G}
\frac{\frac{3}{4}N_f+N_b}{\frac{7}{8}N_f
+ N_b}\left[\frac{r_{init +}^{D-2}}{D-2}+
2(D-3)\lambda r_{init +}^{D-4}\right]$$
where $r_{init +}$ is the initial  horizon radius of a black hole with
mass $M_{init}$  and, interestingly, the  ratio of a given species $i$
to the total emission is given by~:
$$\frac{N_i}{N_{tot}}=\frac{\alpha_s g_i}{\frac{3}{4}N_f+N_{tot}}$$
where $\alpha_s$ is 1 for bosons  and is 3/4 for fermions and $g_i$ is
the  number  of   internal  degrees  of  freedom  for  the  considered
particles. The  mean number of  particles emitted by a Schwarzschild -
Gauss - Bonnet  black hole  ranges from  25  to 4.7  depending on  the
values of $\lambda$ and $D$, for $M_D\sim 1$ TeV and $M_{init}\sim 10$
TeV. Those values are decreased to  5 and 1.05 if $M_{init}$ is set at
2  TeV.  Figure \ref{spec}  shows  the flux  for  different values  of
$\lambda$  and  $D$. Although some combinations seem  to  be  strongly
degenerated, the next  section  shows that in any  case  the values of
$\lambda$ and $D$ can be well reconstructed.

\section{String coupling constant measurement}

\begin{figure}
[htbp]
$$
\epsfxsize=13cm
\epsfysize=12cm
\epsfbox{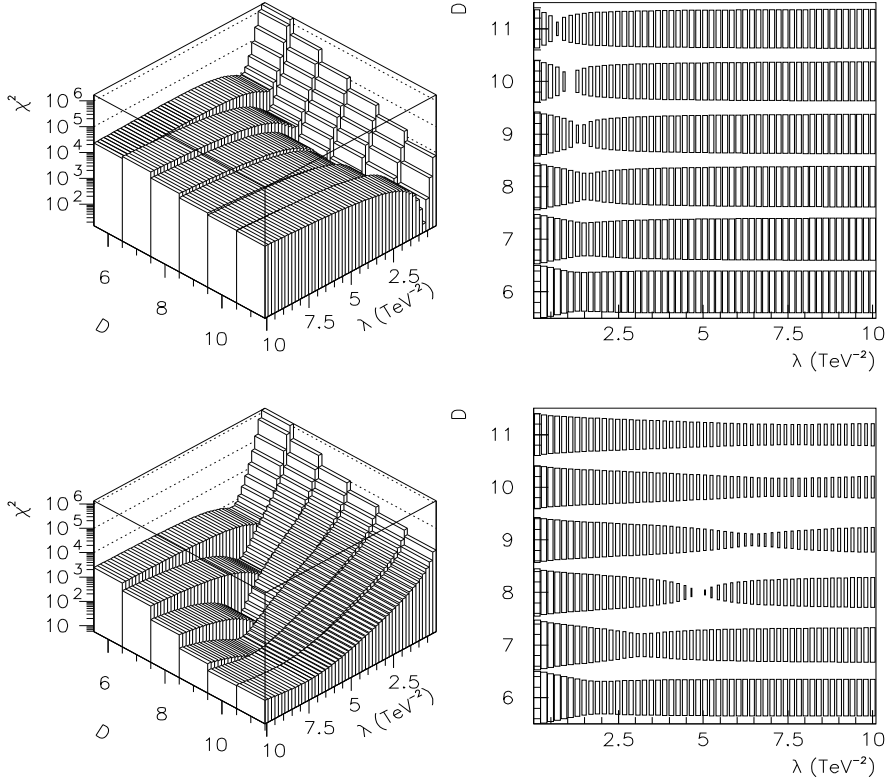}
$$
\caption{Upper part~: values of the $\chi^2/d.o.f.$
for the  reconstructed spectra as a function of  $D$ and $\lambda$ for
"input" values $\lambda=1$~TeV$^{-2}$ and $D=10$ ; the right side shows 
rectangles proportional to the logarithm of the $\chi^2/d.o.f.$ 
Lower part (left and
right)~: values of  the  $\chi^2/d.o.f.$ for the reconstructed spectra
as   a   function   of   $D$   and   $\lambda$   for   "input"  values
$\lambda=5$~TeV$^{-2}$ and $D=8$ ; the right side shows 
rectangles proportional to the logarithm of the $\chi^2/d.o.f.$}
\label{lambda}
\end{figure}

To  investigate  the  LHC  capability to reconstruct  the  fundamental
parameter $\lambda$, we have fixed the Planck scale at 1 TeV. Although
a  small  excursion  range  around   this   value   would  not  change
dramatically our  conclusions, it cannot be  taken much above,  due to
the  very  fast  decrease of the  number  of  formed  black holes with
increasing $M_D$. Following \cite{dimo}, we  consider  the  number  of
black holes  produced between 1 TeV and 10 TeV with a bin width of 500
GeV (much larger than the energy resolution of the detector), rescaled
with the value of $r_+$  modified  by the Gauss - Bonnet term. For  each
black  hole event,  the  emitted particles are  randomly  chosen by  a
Monte-Carlo simulation according  to the spectra given in the previous
section, weighted by the appropriate number of degrees of freedom. The
Hawking radiation takes
place predominantly in the S-wave channel \cite{emparan}, so
bulk modes  can be neglected and the evaporation  can be considered as
occurring within the brane. As the intrinsic spectrum ${\rm d}N_i/{\rm
dQ}$  is  very strongly modified by fragmentation  process,  only  the
direct emission of electrons and photons above 100  GeV is considered.
We  have  checked with the Pythia \cite{pythia} hadronization  program
that  only  a  small  fraction of directly emitted  $\gamma$-rays  and
electrons  fall  within  an  hadronic jet, making them  impossible  to
distinguish  from  the background of decay products. Furthermore,  the
background from  standard  model  Z(ee)+jets and $\gamma$+jets remains
much lower than the expected signal.  The value of the Planck scale is
assumed to be known  as a clear threshold effect should appear  in the
data and a negligible uncertainty is expected on this measurement. For
each event, the initial mass of the black hole is  also  assumed to be
known as it can  be easily determined with the full spectrum  of decay
products  (only  5\% of missing energy  is  expected due to the  small
number of  degrees of freedom  of neutrinos and gravitons). The energy
resolution  of  the detector is taken into  account  and  parametrized
\cite{atlas}    as    $\sigma/E=\sqrt{a^2/E+b^2}$    with    $a\approx
10$\%$\sqrt{{\rm GeV}}$  and  $b\approx 0.5$\%. Unlike \cite{dimo}, we
also take into  account  the  time evolution of the  black  holes  and
perform a  full fit  for each event. Once all  the particles have been
generated,  spectra  are  reconstructed  for  all  the  mass  bins and
compared  with  theoretical  computations.  The  values   of  $D$  and
$\lambda$  compatible  with the simulated data are then  investigated.
Figure \ref{lambda}  shows  the  $\chi^2/d.o.f.$ for the reconstructed
spectra for  2  different couples ($\lambda~ [~TeV^{-2}]$, $D$)=(1,10)
and ($\lambda~ [~TeV^{-2}]$, $D$)=(5,8). The statistical  significance
of this $\chi^2$  should  be  taken with care since  a  real  statistical
analysis would require  a full Monte-Carlo simulation of the detector.
Nevertheless, the "input" values can clearly  be  extracted  from  the
data.  Furthermore,  it  is  important to notice that  for  reasonable
values of  $\lambda$ (around the order  of the quantum  gravity scale,
{\it i.e.} around a TeV$^{-2}$  in  our case) it can unambiguously  be
distinguished between the case {\it with} and the case {\it without} a
Gauss - Bonnet term. Table \ref{tab:val}  summarizes  the LHC
reconstruction capability requiring the $\chi^2/d.o.f.$ to remain smaller
than  $2\chi^2_{min}/d.o.f.$ where $\chi^2_{min}/d.o.f.$ 
corresponds to the
``physical" case ({\it i.e.} $\lambda=\lambda_{input}$ and $D=D_{input}$). 
This is quite conservative and should translate into high confidence levels
which would require a much more detailed modelling of the 
detector to be accurately computed. For each set of parameters, the cross
section has been taken as $\pi r_+^2$, $\pi r_+^2/10$, 
$\pi r_+^2/100$ and $\pi r_+^2/1000$ to account for uncertainties on the 
production process for $D>4$ with a non-zero impact parameter. Based on the
methods developed by Penrose and D'Eath \& Payne \cite{supp} and on the hoop
conjecture \cite{supp2}, several estimates have been derived and confirm 
the formation of an apparent horizon. The wide range investigated in
this study should account for all physical cases.

\begin{table}[htbp]
\begin{center}
\begin{tabular}{|p{2.0cm}|*{4}{c|}|}
\hline
Allowed values (min/max) & $\lambda = 0.5$ TeV$^{-2}$ & $\lambda  = 1$
TeV$^{-2}$ & $\lambda = 5$ TeV$^{-2}$ \\
\hline
$D$=6 &  $\lambda: 0.39/0.58$ ; $D: 6/6$ &  $\lambda: 0.78/1.18$ ; $D:
6/6$ &  $\lambda:  > 3.15 $ ; $D: 6/7$ \\
 & $\lambda: 0.39/0.58$ ; $D:
6/6$ & $\lambda: 0.78/1.18$ ;  $D:  6/6$ & $\lambda: >3.15$ ;  $D:
6/8$ \\
 & $\lambda: 0.39/0.58$ ; $D:
6/6$ & $\lambda: 0.78/1.18$ ;  $D:  6/6$ & $\lambda: >2.20$ ;  $D:
6/8$\\
& $\lambda: 0.39/0.58$ ; $D:
6/6$ & $\lambda: 0.78/1.32$ ;  $D:  6/7$ & reconstruction fails\\
\hline
$D$=7 &  $\lambda: 0.39/0.58$ ; $D: 7/7$ &  $\lambda: 0.78/1.18$ ; $D:
7/7$ &  $\lambda:  > 3.96 $ ; $D: 7/8$ \\
 & $\lambda: 0.39/0.58$ ; $D:
7/7$ & $\lambda: 0.78/1.18$ ;  $D:  7/7$ & $\lambda: >3.77$ ;  $D:
7/9$ \\
 & $\lambda: 0.39/0.58$ ; $D:
7/7$ & $\lambda: 0.78/1.18$ ;  $D:  7/8$ & $\lambda: >3.56$ ;  $D:
7/9$ \\
& $\lambda: 0.16/0.58$ ; $D:
7/8$ & $\lambda: 0.18/1.37$ ;  $D:  7/11$ & $\lambda: >1.58$ ;  $D:
6/11$ \\
\hline
$D$=8 &  $\lambda: 0.39/0.58$ ; $D: 8/8$ &  $\lambda: 0.99/1.18$ ; $D:
8/8$ & $\lambda: 4.56/6.92$ ; $D: 8/9$ \\
 & $\lambda: 0.39/0.58$ ; $D:
8/8$ &  $\lambda: 0.99/1.18$ ; $D: 8/8$ &  $\lambda: 4.34/7.50$ ; $D:
8/9$ \\
 & $\lambda: 0.39/0.58$ ; $D:
8/8$ &  $\lambda: 0.77/1.18$ ; $D: 8/9$ &  $\lambda: >3.95$ ; $D:
8/11$ \\
 & $\lambda: 0.20/0.79$ ; $D:
7/9$ &  $\lambda: 0.22/1.56$ ; $D: 7/11$ &  $\lambda: >2.34$ ; $D:
7/11$ \\
\hline
$D$=9 &  $\lambda: 0.39/0.58$ ; $D: 9/9$ &  $\lambda: 0.99/1.18$ ; $D:
9/9$ & $\lambda: 4.74/5.34$ ; $D: 9/9$ \\ 
& $\lambda: 0.39/0.58$ ; $D:
9/9$ & $\lambda: 0.99/1.18$ ; $D: 9/9$ & $\lambda: 4.55/5.91$  ; $D:
9/10$ \\
& $\lambda: 0.18/0.58$ ; $D:
9/10$ & $\lambda: 0.37/1.18$ ; $D: 9/11$ & $\lambda: 3.59/7.29$  ; $D:
8/11$ \\
& $\lambda: <0.96$ ; $D:
8/11$ & $\lambda: 0.22/1.58$ ; $D: 8/11$ & $\lambda: >2.37$  ; $D:
7/11$ \\
\hline
$D$=10 &  $\lambda: 0.18/0.58$ ; $D:  10/11$ & $\lambda:  0.99/1.18$ ;
$D:  10/10$  &  $\lambda:  4.74/5.53$  ;  $D: 10/10$  \\
  &  $\lambda:
0.18/0.58$ ; $D: 10/11$ & $\lambda: 0.58/1.18$ ; $D: 10/11$  & $\lambda:
4.36/5.71$ ; $D: 10/11$ \\
&  $\lambda:
0.18/0.58$ ; $D: 10/11$ & $\lambda: 0.58/1.58$ ; $D: 9/11$  & $\lambda:
3.58/6.72$ ; $D: 9/11$ \\
&  $\lambda:
0.18/0.97$ ; $D: 9/11$ & $\lambda: 0.39/1.96$ ; $D: 8/11$  & $\lambda: >2.77$ 
; $D: 8/11$ \\
\hline
$D$=11 &  $\lambda: 0.39/0.99$ ; $D:  10/11$ & $\lambda:  0.99/1.58$ ;
$D:   10/11$   &    $\lambda:   4.74/5.53$   ;   $D:   11/11$   \\
   &
$\lambda:0.39/0.99$ ;  $D: 10/11$ &  $\lambda: 0.98/1.58$ ; $D: 10/11$ &
$\lambda: 4.57/6.12$ ; $D: 10/11$ \\
&
$\lambda:0.39/0.99$ ;  $D: 10/11$ &  $\lambda: 0.75/1.77$ ; $D: 10/11$ &
$\lambda: 4.14/7.16$ ; $D: 9/11$ \\
&
$\lambda:0.39/1.56$ ;  $D: 9/11$ &  $\lambda: 0.75/2.37$ ; $D: 9/11$ &
$\lambda: >2.96$ ; $D: 8/11$ \\
\hline
\end{tabular}
\end{center}
\caption{Reconstructed values for $D$  and  $\lambda$~(TeV$^{-2}$) as a function of
the "real" input values requiring  $\chi^2<2\chi^2_{min}$. The first line assumes
$\sigma=\pi r_+^2$, the second line $\sigma=\pi r_+^2/10$, 
the third line $\sigma=\pi r_+^2/100$ and the fourth line $\sigma=\pi
r_+^2/1000$.}
\label{tab:val}
\end{table}

\section{Discussion}

In  case  the Planck  scale  lies  in  the  TeV  range  due  to  extra
dimensions, this study shows that, beyond the dimensionality of space,
the  next  generation of  colliders  should  be  able  to  measure the
coefficient  of  a  possible  Gauss - Bonnet term in  the  gravitational
action. This would allow an important step forward in the construction
of a full quantum theory of  gravity. It is also interesting to notice
that  this  would  be  a  nice  example  of  the  convergence  between
astrophysics and particle  physics in the final understanding of black
holes and gravity in the Planckian region.

Nevertheless, those results could  be  improved and refined in several
ways. First, the endpoint of the Hawking evaporation  process is still
an unsolved  problem. In this paper, we have  considered that the time
integral  of  the instantaneous  spectrum  is valid  up  to the  total
disappearance of the black hole. Although usually a good approximation
(as most  particles are emitted at masses close  to the initial mass),
this can become a serious problem if the number of extra dimensions is
high. In such cases, the mean  number of emitted particles can be very
small and even smaller than one. The spectrum therefore {\it  must} be
truncated properly. A possibility could be to add a Heavyside function
to  ensure  energy  conservation  while keeping the  same  probability
distribution,  as   suggested   in   \cite{stasastro},   but   a  full
understanding of the phenomenon would be required as the analytical
formulae derived in this work would not stand anymore.

Then, as studied in \cite{cai,bir}, a cosmological constant could also
be  included  in the action. On  the  theoretical side, this would  be
strongly motivated by the great deal of attention paid to  the Anti-de
Sitter and,  recently, de Sitter / Conformal Field  Theory (AdS and dS
/CFT) correspondences. On the experimental  side,  this  would open an
interesting window as there is no  unambiguous  relation  between  the
D-dimensional and the 4-dimensional cosmological constants.

Finally, it would be  very interesting to extend this study to  Kerr -
Gauss  -  Bonnet  black  holes \cite{aurelnew} as the  holes  possibly
produced   at   colliders  are  expected  to  be  spinning.   Although
qualitatively   equivalent,   the   results   are   expected   to   be
quantitatively quite different and probably more realistic.

\section*{Acknowledgments}

S.A.  would like to  thank  the  AMS  Group in  the  ``Laboratoire  de
Physique Subatomique et de Cosmologie  (CNRS/UJF)  de  Grenoble''  for
kind hospitality during  the  first part of this  work.  This work was
also  supported   (S.A.)  by  ``Universities  of  Russia:  Fundamental
Investigations'' via grant No. UR.02.01.026.


\begin{thebibliography}{99}

\bibitem{dimo}
S. Dimopoulos \& G. Landsberg, Phys. Rev. Lett. 87 (2001) 161602

\bibitem{gidd}
S.B. Giddings \& S. Thomas, Phys. Rev. D 65 (2002) 056010

\bibitem{idea}
K. Cheung, Phys. Rev. Lett. 88 (2002) 221602\\
P. Kanti \& J. March-Russell, Phys. Rev. D 66 (2002) 024023\\
A.V. Kotwal \& C.Hays, Phys. Rev. D 66 (2002) 116005\\
S. Hossenfelder, S.  Hofmann, M. Bleicher  \& H. Stocker, Phys. Rev. D
66 (2002) 101502\\
A. Chamblin \& G.C. Nayak, Phys. Rev. D 66 (2002) 091901\\
V. Frolov \& D. Stojkovic, Phys. Rev. D 66 (2002) 084002\\
M. Cavaglia, Phys. Lett.B 569 (2003) 7-13\\
D. Ida, K.-Y. Oda \& S.C. Park, Phys. Rev. D 67 (2003) 064025\\
M. Cavaglia, S. Das  \& R. Maartens,  Class. Quantum Grav.  20  (2003)
L205\\
R. Casadio \& B. Harms, Int. J. Mod. Phys. A 17 (2002) 4635\\
P. Kanti \& J. March-Russell, Phys. Rev. D 67 (2003) 104019\\
I.P. Neupane, Phys. Rev. D 67 (2003) 061501

\bibitem{ancho}
A. Ringwald \& H. Tu, Phys. Lett. B 525 (2002) 135\\
R. Emparan, M. Masip \& R. Rattazzi, Phys. rev. D 65 (2002) 064023\\
J.L. Feng \& A.D. Shapere, Phys. Rev. Lett. 88 (2002) 021303\\
L.A. Anchordoqui, J.L. Feng, H. Goldberg \& A.D. Shapere, Phys.
Rev. D 65 (2002) 124027\\
E.-J. Ahn, M. Ave, M. Cavaglia  \& A.V. Olinto, Phys. Rev. D 68 (2003)
043004

\bibitem{myers}
R.C. Myers \& M.J. Perry, Ann. Phys. (N.Y.) 172 (1986) 304

\bibitem{large}
N. Arkani-Hamed, S. Dimopoulos \& G.R. Dvali, Phys. Lett. B 429 (1998)
257\\
I. Antoniadis {\it et al.}, Phys. Lett. B 436 (1998) 257\\
N. Arkani-Hamed, S.  Dimopoulos \& G.R.  Dvali, Phys. Rev. D 59 (1999)
086004

\bibitem{rs}
L. Randall \& R. Sundrum, Phys. Rev. Lett. 83 (1999) 3370

\bibitem{katz}
S.B. Giddings\& E. Katz, J. Math. Phys. 42 (2001) 3082

\bibitem{supp}
D.M. Eardley \& S.B. Giddings, Phys. Rev. D 66 (2002) 044011

\bibitem{supp2}
H. Yoshino \& Y. Nambu, Phys. Rev. D 66 (2002) 065004

\bibitem{boulw}
D.G. Boulware \& S. Deser, Phys. Rev. Lett. 55 (1985) 2656

\bibitem{zw}
B. Zwiebach, Phys. Lett. B 156 (1985) 315\\
N. Deruelle \& J. Madore, Mod. Phys. Lett. A 1 (1986) 237\\
N, Deruelle \& L. Farina-Busto, Phys. Rev. D 41 (1990) 3696

\bibitem{mavro}
S. Nojiri, S.D. Odintsov \& S. Ogushi, Phys. Rev. D 65 (2002) 023521\\
M.E. Mavrotamos \& J. Rizos, Phys. Rev. D 62 (2000) 124004\\
Y.M. Cho, I.P. Neupane and P.S. Wesson, Nucl. Phys. B 621 (2002) 388

\bibitem{cosmo}
B.C. Paul \& S. Mukherjee, Phys. Rev. D 42 (1990) 2595\\
B. Abdesselam  \& N. Mohammedi, Phys. Rev. D 65 (2002) 084018\\
C. Charmousis \& J.-F. Dufaux, Class. Quantum .Grav. 19 (2002) 4671\\
J.E. Lidsey \& N.J. Nunes, Phys. Rev. D 67 (2003) 103510

\bibitem{bh}
S.O. Alexeyev \& M.V. Pomazanov, Phys. Rev. D 55 (1997) 2110\\
S.O. Alexeyev, A.  Barrau, G. Boudoul,  O. Khovanskaya \&  M.  Sazhin,
Class. Quantum Grav. 19 (2002) 4431\\
M. Banados, C.  Teitelboim \& J. Zanelli,  Phys. Rev. Lett.  72 (1994)
957\\
T. Torii \& K.-I. Maeda, Phys. Rev. D 58 (1998) 084004

\bibitem{cai}
R.-G. Cai, Phys. Rev. D 65 (2002) 084014\\
A. Padilla, Class. Quantum Grav. 20 (2003) 3129

\bibitem{kanti}
C. M. Harris \& P. Kanti, JHEP 010 (2003) 14

\bibitem{barrau}
A. Barrau {\it et al.}, Astronom. Astrophys. 388 (2002) 676

\bibitem{emparan}
R. Emparan, G. T. Horowitz and R.  C. Myers, Phys. Rev. Lett 85 (2000)
499

\bibitem{pythia}
T. Tj\"ostrand, Comput. Phys. Commun., 82 (1994) 74

\bibitem{atlas}
ATLAS TDR 14, vol 1 CERN/LHCC/99-14 (1999)

\bibitem{stasastro}
S. O. Alexeyev, A. Barrau, G. Boudoul, M. Sazhin \& O. S. Khovanskaya,
Astronom. Letters 38, 7 (2002) 428

\bibitem{bir}
D. Birmingham, Class. Quantum Grav. 16 (1999) 1197

\bibitem{aurelnew}
S. Alexeyev, N. Popov, A. Barrau \& J. Grain, in preparation (2003)

\end{thebibliography}
\end{document}